\documentclass[prb,reprint,superscriptaddress,showpacs,amsmath,amssymb,floatfix,
]{revtex4-1}
\usepackage{dcolumn}
\usepackage{bm}
\usepackage{turnstile}
\usepackage{graphicx}
\usepackage{amsmath,amssymb}
\usepackage{url}
\usepackage{multirow}
\usepackage{float}
\usepackage[toc,page]{appendix}
\usepackage{color}

\begin{document}
\title{Phase diagrams of the superconducting diode effect in topological hybrid structures}

\author{T.~Karabassov}\email{tkarabasov@hse.ru}
	\affiliation{HSE University, 101000 Moscow, Russia}
\author{I.~V.~Bobkova}
 \affiliation{Institute of Solid State Physics, Chernogolovka, Moscow reg., 142432 Russia}
    \affiliation{Moscow Institute of Physics and Technology, Dolgoprudny, 141700 Russia}
    \affiliation{HSE University, 101000 Moscow, Russia}
   \author{V.~M.~Silkin}
   \affiliation{Donostia International Physics Center (DIPC), San Sebasti\'{a}n/Donostia, 20018 Basque Country, Spain}
\affiliation{Departamento de F\'{\i}sica de Materiales, Facultad de Ciencias Qu\'{\i}micas,
    UPV/EHU, 20080 San Sebasti\'{a}n, Basque Country, Spain}
\affiliation{IKERBASQUE, Basque Foundation for Science, 48011 Bilbao, Spain}
\author{.~G.~Lvov}
\affiliation{HSE University, 101000 Moscow, Russia}

\author{A.~A.~Golubov}
\affiliation{Faculty of Science and Technology and MESA$^+$ Institute for Nanotechnology,
		University of Twente, 7500 AE Enschede, The Netherlands}
\author{A.~S.~Vasenko}
    \affiliation{HSE University, 101000 Moscow, Russia}
    \affiliation{I.E. Tamm Department of Theoretical Physics, P.N. Lebedev Physical Institute, Russian Academy of Sciences, 119991 Moscow, Russia}
\begin{abstract}
Recently the superconducting diode effect (SDE) has attracted a lot of attention due to new possibilities in the field of superconducting electronics. One of the possible realizations of the SDE is the implementation in superconducting hybrid structures. In this case the SDE is achieved by means of the proximity effect. However, the optimal conditions for the SDE quality factor in hybrid devices remain unclear. In this study we consider the Superconductor/Ferromagnet/Topological insulator (S/F/TI) hybrid device and investigate the diode quality factor at different parameters of the hybrid structure. Consequently, we reveal important parameters that have crucial impact on the magnitude of the SDE quality factor. 
\end{abstract}
\maketitle

\section{Introduction}
 The superconducting (or supercurrent) diode effect (SDE) is the superconducting analog of the semiconducting diode effect in $p-n$ junctions. While in normal systems the diode effect corresponds to the conduction of the normal current in the only one direction, the superconducting diode effect involves the nonreciprocity of the supercurrent \cite{Nadeem_arxiv}. The discovery of such effect brings many potential applications in low-power logic circuits as quantum computing and spin-based electronics \cite{Eschrig2015,Linder2015}.
 
Various systems that can behave as superconducting diodes have been recently theoretically proposed \cite{Daido2022,He_arxiv,Yuan_arxiv,Scammell_arxiv,Ilic_arxiv,Devizorova2021,dePicoli_arxiv,Grein2009,Lu_arxiv,Legg2022_arxiv}  and experimentally discovered \cite{Ando2020,Bauriedl_arxiv,Shin_arxiv,Trahms_arxiv, Chahid2023, Chahid_arxiv,Suri2022,Hou_arxiv,Hideki_arxiv,Lyu2021}. The superconducting diode effect can be realized in two-dimensional (2D) superconducting systems if both inversion and time-reversal symmetries are broken\cite{Nadeem_arxiv}. One of the most promising SDE devices are Josephson diodes, where weak link plays the key role in achieving the current nonreciprocity \cite{Bocquillon2017,Baumgartner2022,Wu2022,Pal_arxiv,Baumgartner_arxiv,Zhang_arxiv,Hu2007,Chen2018,Yokoyama2014,Kopasov2021,Davydova_arxiv,Halterman2022,Alidoust2021,Tanaka2022_arxiv,Golod2022_arxiv,Kokkeler2022}. As it has been shown recently the SDE in Josephson junctions may be due to the only inversion symmetry breaking and do not require the time-reversal symmetry breaking \cite{Zhang_arxiv}.  It can be implemented in the weak links with voltage dependent Rashba spin-orbit coupling (SOC) or electric polarization, which leads to $I_c(V) \ne I_c(-V)$, where $I_c$ is the Josephson critical current. For example, the field-free inversion asymmetrical Josephson diode has been reported recently\cite{Wu2022}.

The ``conventional'' SDE devices require both inversion and time-reversal symmetry breaking in 2D superconducting films \cite{Nadeem_arxiv}. The inversion symmetry can be broken by introducing the spin-orbit coupling. Experimentally it can be achieved in hybrid structures by proximity to a three-dimensional topological insulator (TI), in superconductors with Rashba spin-orbit coupling (like in polar SrTiO$_3$ films\cite{Itahashi2020}, few-layer MoTe$_2$ in the $T_d$ phase or MoS$_2$ \cite{Yuan_arxiv,Wakatsuki2017,Wakatsuki2018}, and twisted bilayer graphene\cite{Lin_arxiv,Efetov2023}), or, in some cases, by the asymmetry of the device geometry. On the other hand, the time-reversal symmetry can be broken by the in-plane magnetic field, or alternatively in hybrid structures by proximity to a ferromagnetic insulator with the in-plane exchange field. Such conditions can lead to $I_c^+ \ne I_c^-$ (Here $I_c^+$  and $I_c^-$ are the critical currents in the opposite directions). 

The breaking of inversion and time-reversal symmetries in 2D superconductors allows for a formation of the helical superconducting phase \cite{Edelstein1989,Barzykin2002,Dimitrova2007,Samokhin2004,Kaur2005,Houzet2015}, with the order parameter modulated in the direction transverse to the magnetic (or exchange) in-plane field: $\Delta({\bm{r}}) = \Delta \exp\left( i {\bm{q_0 r}} \right)$. Its physical origin can be explained as follows. The SOC produces the spin-momentum locking term $\propto \bm n \cdot (\bm \sigma \times \bm p)$ in the Hamiltonian, where $\bm n$ is the unit vector  perpendicular to the plane of the system, $\bm p$ is the electron momentum and $\bm \sigma$ is its spin. The applied field makes spin-down state energetically more favorable. Due to the spin-momentum locking it results in the fact that one of the mutually opposite momentum directions along the axis perpendicular to the field direction is more favorable, which leads to the appearance of the spontaneous current. However, the superconductor develops a phase gradient, which exactly compensates the spontaneous current. The resulting phase-inhomogeneous zero-current state with finite $\bm{q_0}$ is the true ground state of the system. This state looks similar to another well-known inhomogeneous superconducting state, FFLO state \cite{Fulde1964,Larkin1965,Mironov2012,Mironov2018}.

The situation when the exchange field, superconductivity and SOC coexist intrinsically is rare and largely unexplored from the point of view of magnetoelectrics \cite{Bobkova2016}. At the same time bringing all ingredients together can be easily achieved in hybrid devices based on a combination of conventional materials: superconductors (S), three-dimensional topological insulators (TI) and ferromagnet (F). Hybrid superconducting diode can be realized as a F/S/TI/S/F Josephson junction (Josephson SDE) \cite{Kokkeler2022}, or a S/F bilayer on top of a three-dimensional topological insulator (proximity SDE) \cite{Karabassov2021,Karabassov2022}. In this paper we consider the latter case. TI is chosen because its conductive surface state exhibits full spin-momentum locking due to a very strong SOC: an electron spin always makes a right angle with its momentum \cite{Burkov2010,Culcer2010,Yazyev2010,Li2014}. It was found that although the exchange field and superconducting order parameter are spatially separated, the finite-momentum helical state is realized, accompanied by the spontaneous currents parallel to the S/F interface, inhomogeneously distributed over the bilayer in such a way that the net current vanishes \cite{Karabassov2022}. Such a hybrid state only takes place when the exchange field has a component perpendicular to the S/F interface. This hybrid state is intrinsically nonreciprocal and can be used as a superconducting diode. If a current parrallel to the S/F interface flows in such a structure, in one direction it is a supercurrent, whereas in other direction it can be a dissipative current.

The nonreciprocity of the critical current can be quantified by the superconducting diode quality factor,
\begin{equation}
\eta = \frac{I_{c}^+ - I_{c}^-}{I_{c}^+ + I_{c}^-},
\end{equation}\label{eta}
where the $I_{c}^\pm$ are external positive and negative currents parallel to the S/FI interface. We have already shown that $\eta$ is finite in the considered system \cite{Karabassov2022}. However, the optimal conditions for the SDE quality factor remain unclear. In this paper we investigate the diode quality factor of the hybrid structure, studying the phase diagrams of the superconducting diode effect for various parameters of the S/F/TI hybrid structure which influence the superconducting diode effect. Our consideration is based on the microscopic quasiclassical theory of superconductivity in terms of the Usadel equations. This study can be useful for the experimental fabrication of the superconducting diode with sufficiently good efficiency for applications in superconducting electronics.

The paper is organized as follows. In Sec.~\ref{Model} we formulate theory of the superconducting diode effect in the framework of the quasiclassical Usadel equations for mesoscopic superconductors. In Sec.~\ref{Results} we use this theory to calculate the phase diagrams of the superconducting diode effect for various parameters of the hybrid structure under consideration. Finally, we summarize the key points of the research in Sec.~\ref{Fin}.

\section{Model}\label{Model}

\begin{figure}[t]
	\includegraphics[width=\columnwidth,keepaspectratio]{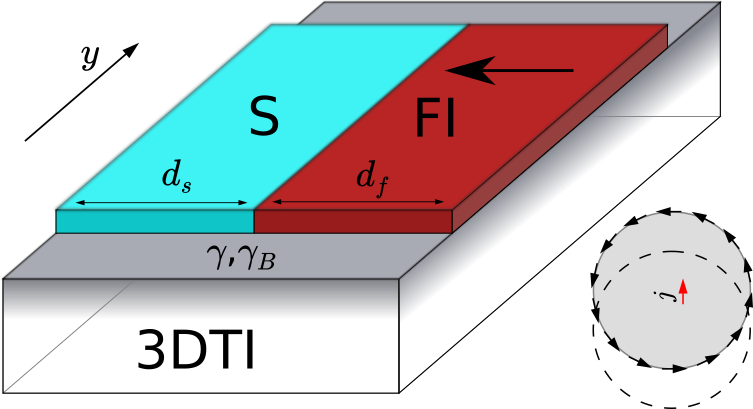}
	\caption{ (a)  Schematic geometry of the S/F (Ferromagnetic insulator) bilayer on top of the 3D TI. Right bottom corner: Fermi-surface of the TI surface states and the illustration of the magnetolectric effect. The quasiparticle spin $\bm S$ is locked at the right angle to its momentum $\bm p$.
	}
	\label{model}
\end{figure}
In this section we present the system under consideration depicted in Fig. \ref{model}.  It consists of a thin superconductor (S) deposited on top of the TI and adjacent ferromagnetic insulator (F). This system is described by the following model Hamiltonian:
\begin{equation}
H= H_0 + H_F + H_{S},
\end{equation}
where
\begin{eqnarray}
H_0=\int d^2 r \Psi^\dagger (\bm r)\bigl[-i\alpha(\bm \nabla_{\bm r}\times \hat z)\bm \sigma - \mu +  V(\bm r) \bigr]\Psi(\bm r), \\
H_F = - \int d^2 r \Psi^\dagger (\bm r) \bigl[ \bm h \bm \sigma \bigr] \Psi(\bm r), \\
H_S = \Delta(\bm r)\Psi^\dagger_\uparrow (\bm r) \Psi^\dagger_\downarrow (\bm r) + \Delta^*(\bm r)\Psi_\downarrow (\bm r) \Psi_\uparrow (\bm r).
\end{eqnarray}
Here $\Psi^\dagger(\bm r)=(\Psi^\dagger_\uparrow(\bm r),\Psi^\dagger_\downarrow(\bm r))$ is the creation operator of an electron at the 3D TI surface, $\hat z$ is the unit vector normal to the surface of TI, $\alpha$ is the Fermi velocity of electrons at the 3D TI surface and $\mu$ is the chemical potential. $\bm \sigma = (\sigma_x, \sigma_y, \sigma_z)$ is a vector of Pauli matrices in spin space and $\bm h = (h_x, h_y, 0)$ is an in-plane exchange field, which is assumed to be nonzero only at $x<0$. The superconducting pairing potential $\Delta $ is nonzero only  at $x>0$. Therefore, effectively the TI surface states are divided into two parts: one of them at $x<0$ possesses $h \neq 0$ and can be called "ferromagnetic", while the other part corresponding to $x>0$ with $\Delta \neq 0$ can be called "superconducting". Below we will use subscripts $f$ and $s$ to denote quantities, related to the appropriate parts of the TI surface. The potential term $V(\bm r)$ includes the nonmagnetic impurity scattering potential $V_{imp}=\sum \limits_{\bm r_i}V_i \delta(\bm r - \bm r_i)$, which is of a Gaussian form $\langle V(\bm r)V(\bm r')\rangle = (1/\pi \nu \tau)\delta(\bm r - \bm r')$ with $\nu=\mu/(2\pi \alpha^2)$.

The superconductivity and in-plane exchange field is assumed to be proximity induced due to adjacent superconducting and ferromagnetic layers. Thus we can imagine the system to be a planar hybrid structure that consists of superconductor S and ferromagnetic layer F on top of three-dimensional topological insulator TI  as shown schematically in Fig.~\ref{model}. The role of the TI surface is to provide a strong spin-orbit coupling which produces a full spin-momentum locking effect. In this case only one helical band which crossing the Fermi energy is present. We employ the quasiclassical Green's function formalism in the diffusive regime. In principle Green's function matrices have two degrees of freedom that are particle-hole and spin. In our model the spin structure is characterized by a projector onto the conduction band:
\begin{equation}
\check g_{s,f}(\bm n_F, \bm r, \varepsilon)= \hat g_{s,f}(\bm r, \varepsilon)\frac{(1+\bm n_\perp \bm \sigma)}{2},
\label{spin_structure}
\end{equation}
where $\hat g_{s(f)}$ is the spinless Green's functions matrix in the particle-hole space in the superconducting (ferromagnetic) part of the 3D TI layer, $\bm n_F=\bm p_F/p_F=(n_{F,x},n_{F,y},0)$ is a unit vector directed along the quasiparticle trajectory and $\bm n_\perp=(n_{F,y},-n_{F,x},0)$ is a unit vector perpendicular to the quasiparticle trajectory and directed along the quasiparticle spin, which is locked to the quasiparticle momentum. 

In our theoretical analysis, we consider the diffusive limit, in which the superconducting coherence length is given by expression $\xi_{s} = \sqrt{D_{s}/ 2 \pi T_{cs}}$, where $D_{s}$ is the diffusion coefficient and $T_{cs}$ is the critical temperature of the bulk superconductor (we assume $\hbar = k_B = 1$) and the elastic scattering length $\ell \ll \xi_{s}$. We also neglect the nonequilibrium effects in the structure \cite{VH}.

In the following we outline the nonlinear equations to calculate the SDE effect in the system under consideration. The quasiclassical Usadel equation for spinless Green's functions is\cite{Zyuzin2016,Bobkova2017}
\begin{equation}\label{Usadel_general}
D \hat{\nabla}\left(\hat{g} \hat{\nabla} \hat{g} \right)= \left[\omega_n \tau_z + i \hat{\Delta}, \hat{g}\right].
\end{equation}
Here $D$ is the diffusion constant, $\tau_z$ is the Pauli matrix in the particle-hole space, $\hat{\nabla} X = \nabla X + i \left(h_x \hat{e}_y - h_y \hat{e}_x\right) \left[\tau_z, \hat{g}\right]/\alpha$. The gap matrix $\hat{\Delta}$ is defined as $\hat{\Delta}= \hat{U} i \tau_x \Delta (x) \hat{U}^\dagger$, where $\Delta(x)$ is a real function and transformation matrix $\hat{U}= \exp \left( i q y \tau_z/2 \right)$ . The finite center of mass momentum $q$ takes into account the helical state. The Green's function matrix is also transformed as $\hat{g}= \hat{U} \hat{g}_q \hat{U}^\dagger$.
Inclusion of the magnetization component $h_y$ produces no quantitative effect neither on the supercurrent in $y$ direction of the bilayer nor on the critical temperature in the S part. It only enters the solution $f_f$ as a phase factor $\exp\left(2 i h_y x/\alpha\right)$ \cite{Zyuzin2016,Karabassov2021}. Thus we do not take it into consideration in our model and define $h_x=h$.

In the hybrid structure under consideration the helical state appears in the system in the following way. In our system the Zeeman field and superconducting region are spatially separated so that the helical state is realized via the proximity effect through the S/F interface. The helical state in the system is also characterized by the order parameter with spatially inhomogeneous phase, however the supercurrent density is not uniform in the structure. The total current across the hybrid structure is equal to zero. Thus, this state is called hybrid helical state \cite{Karabassov2022}.

To facilitate the solution procedures of the nonlinear Usadel equations we employ $\theta$ parametrization of the Green's functions\cite{Belzig1999},
\begin{equation}
\hat{g}_q= 
\begin{pmatrix}
\cos {\theta} & \sin \theta \\
\sin \theta & -\cos \theta
\end{pmatrix}.
\end{equation}
Substituting the above matrix into the Usadel equation \eqref{Usadel_general}, we obtain in the S part of the TI surface $x>0$:
\begin{equation}
\xi_{s}^2 \pi T_{cs} \left[ \partial_x^2 \theta_{s} - \frac{q^2 }{2} \sin 2 \theta_s \right] =\omega_n \sin{\theta_{s}} - \Delta (x) \cos{\theta_{s}}, \nonumber
\end{equation}
and in the F part $x<0$:
\begin{equation}
\xi_{f}^2 \pi T_{cs}  \left[ \partial_x^2 \theta_{f} - \frac{q_m^2 }{2} \sin 2 \theta_f \right] =\omega_n \sin{\theta_{f}},
\end{equation}
where $\xi_{f} = \sqrt{D_{f}/ 2 \pi T_{cs}}$, and $D_f$ is the diffusion coefficient of the ferromagnetic layer. $q_m = q + 2 h /\alpha$ and $X_{s(f)}$ means the value of $X$ in the S(F) part of the TI surface, respectively. The self-consistency equation for the pair potential reads,
\begin{equation}
\Delta (x) \ln \frac{T_{cs}}{T} = \pi T \sum_{\omega_n} \left( \frac{\Delta (x)}{|\omega_n|} - 2 \sin \theta_s \right).
\end{equation}
We supplement the above equations with the following boundary conditions at the S/F interface ($x=0)$ \cite{KL},
\begin{eqnarray}\label{BC}
\gamma_B \frac{\partial \theta_f }{\partial x}\Big\vert_{x=0} = \sin \left( \theta_s - \theta_f \right),\\
\frac{\gamma_B}{\gamma} \frac{\partial \theta_s}{\partial x}\Big\vert_{x=0} = \sin \left( \theta_s - \theta_f \right),
\end{eqnarray}
where $\gamma = \xi_s \sigma_f/ \xi_f \sigma_s$, $\gamma_B = R \sigma_f / \xi_f$, and $\sigma_{s(f)}$ is the conductivity of the S (F)layer. 
The parameter $\gamma$ determines the strength
of suppression of superconductivity in the S lead near
the interface compared to the bulk: no suppression occurs
for $\gamma = 0$, while strong suppression takes place for $\gamma \gg 1$. While for identical materials $\gamma = 1$, in general this parameter may have arbitrary value.  
The parameter $\gamma_B$ is the dimensionless parameter, describing the 
transparency of the S/F interface \cite{KL, VB1, VB2}. While for an ideal fully transparent interface $\gamma_B=0$, in general case one may expect finite value of $\gamma_B$. In this work we will study the dependence of the diode efficiency on the values of $\gamma$ and $\gamma_B$ in order to find the most favorable conditions for the SDE.

To complete the boundary problem we also set boundary
conditions at free edges,
\begin{equation}\label{BC_vac}
\frac{\partial \theta_f }{\partial x}\Big\vert_{x=-d_f}=0, \quad \frac{\partial \theta_s }{\partial x}\Big\vert_{x=d_s}=0.
\end{equation}

In order to calculate the superconducting current we utilize the expression for the supercurrent density
\begin{equation}
\textbf{J}_{s(f)}= \frac{- i \pi \sigma_{s(f)}}{4 e} T \sum_{\omega_n} Tr \left[ \tau_z \hat{g}_{s(f)} \hat{\nabla} \hat{g}_{s(f)} \right].
\end{equation}
Performing the unitary transformation $U$, the current density transforms as follows:
\begin{eqnarray}
{j}_y^s (x)=- \frac{\pi \sigma_s q }{2 e} T \sum_{\omega_n} \sin^2 \theta_s, \\
{j}_y^f (x)=- \frac{\pi \sigma_n }{2 e} \left[ q  + \frac{2 h}{\alpha}\right] T  \sum_{\omega_n} \sin^2 \theta_f.
\end{eqnarray}
The total supercurrent flowing via the system along the $y$-direction can be calculated by integrated the current density of the total width of the S/F bilayer $d_f+d_s$:
\begin{equation}\label{I_total}
I= \int_{-d_f}^{0} {j}^f_y (x) dx  + \int_{0}^{d_s} {j}^s_y (x) dx .
\end{equation}
In the next section we present the phase diagrams of the quality factor defined in Eq. (\ref{eta}).  In order to compute $\eta$ we find the critical supercurrents $I_c^{+(-)}$ by calculating the total supercurrent as a function of $q$ and finding the maximum and minimum values of $I(q)$. 

\section{Results}\label{Results}

In the present section we introduce the results of the calculations. Here we use $\xi_f = \xi_s$ and dimensionless exchange field $H= \xi h/\alpha$.

\begin{figure}[t]
	\includegraphics[width=\columnwidth,keepaspectratio]{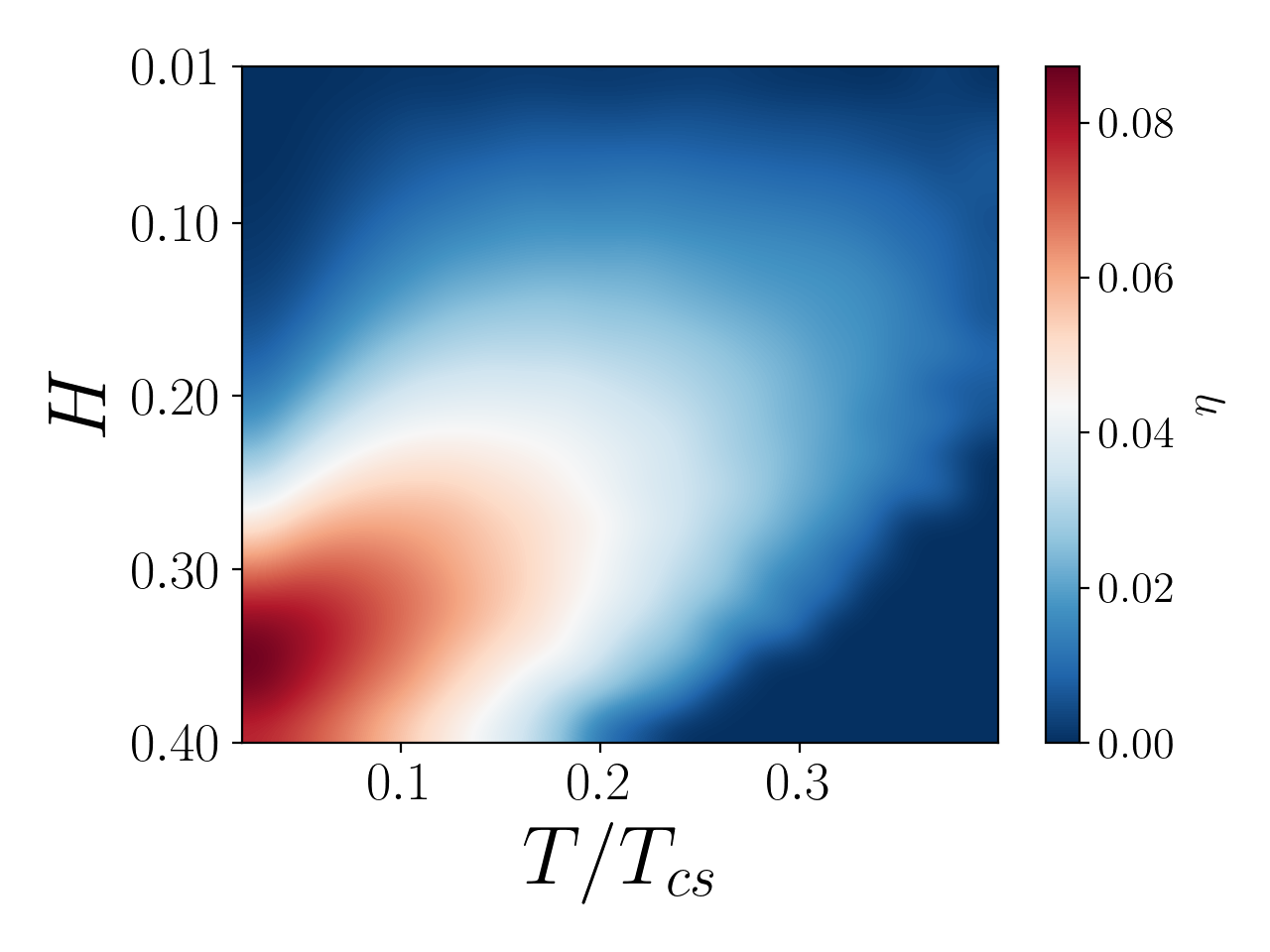}
	\caption{  $H-T$ phase diagram of the diode efficiency $\eta$. The parameters of the calculation: $\gamma = 0.5, \gamma_B=0.3, d_s = 1.2 \xi, d_f = \xi$.
	}
	\label{eta_th}
\end{figure}

In Fig. \ref{eta_th} the dependence of the SDE quality factor $\eta$ as a function of temperature $T_c$ and exchange field $H$ is demonstrated. The exchange field in the considered system can be changed by rotating the magnetization of the adjacent F part. Since the magnetization component along the interface does not introduce any quantitative effect on the observable quantities, the projection onto the $x$ axis $h_x$ ( component normal to the interface ) of the in-plane magnetization $\bm{h}=h_0(\cos \phi, \sin \phi, 0)$ ( $\phi$ is the angle between $\bm{h}$ and $x$ axis) can be varied by rotating the magnetization angle $\phi$. From Fig. \ref{eta_th}  we can notice several important observations. First, we can see that $\eta$ has the largest value at a low temperature. For given parameters the maximum value of $\eta$ corresponds to $H \approx 0.35$. From the figure we can observe that the quality factor decreases as the temperature increased, which is in agreement with the previous studies in other systems\cite{Ilic_arxiv, He_arxiv,Daido2022,Kokkeler2022}. The boundary that corresponds to the vanishing of $\eta$ determines the critical exchange field $H_c$ that destroys the superconducting state in the system. It is unlikely that this boundary may correspond to the $\eta =0$  in the superconducting state. This can be confirmed by simple analytical calculations in the vicinity of the critical temperature.  For instance, in the limit of $ d_f \ll d_s$ and $H d_f/\xi \ll 1 $ the quality factor can be written as \cite{Karabassov2022}
\begin{equation}
    \eta \approx \frac{1}{2}  \frac{\sqrt{7 \zeta (2) \zeta (3) }}{\left(T/T_{cs}\right)^{5/2}} \frac{H d_f}{d_s} \approx 1.86 \frac{1}{\left(T/T_{cs}\right)^{5/2}} \frac{H d_f}{d_s}.
\end{equation}
Here we can see that the quality factor is governed by the parameters of the F part, $d_f$ and $H$. From this expression we can see that as long as the exchange field $H$ and the width $d_f$ are finite the quality factor is distinct from zero. These parameters are non zero in our case, hence the SDE should be present even at the vicinity of the critical temperature $T_c$ . For this reason it is safe to note that this boundary corresponds to the critical field $H_c$. 

\begin{figure}[t]
	\includegraphics[width=\columnwidth,keepaspectratio]{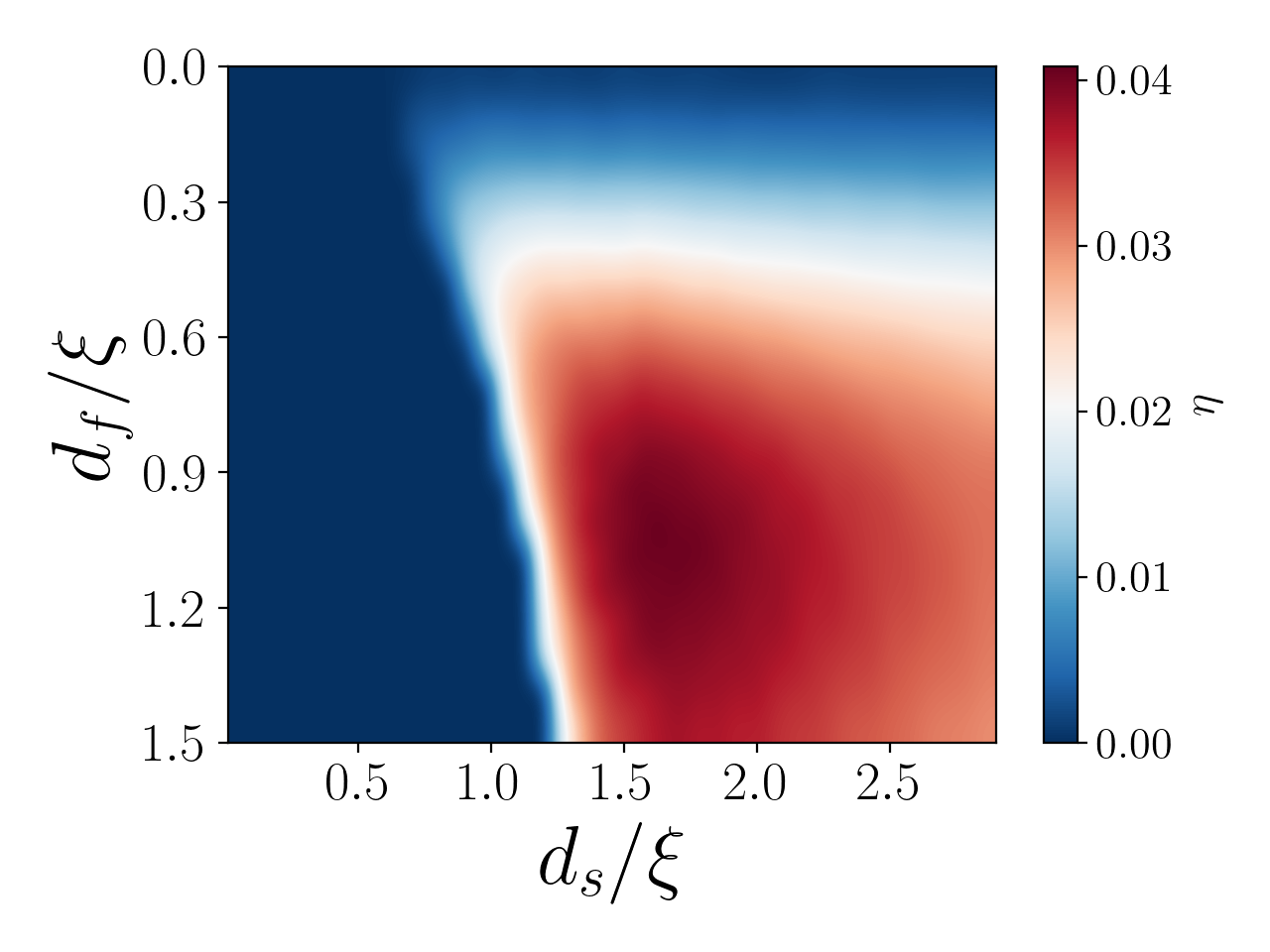}
	\caption{Phase diagram of the diode efficiency $\eta$ plotted on the axes $d_s - d_f$ . The parameters of the calculation: $\gamma = 0.5, \gamma_B=0.3, T = 0.25 T_{cs}, H = 0.3$.
	}
	\label{eta_dsdf}
\end{figure}

It is important to note that in the system under consideration it is impossible to observe the sign change of the SDE. i. e. when $\eta<0$ \cite{Ilic_arxiv,Daido2022}. We consider the system in which the spin structure of the correlations is projected onto a single helical band. In this case we do not take into account the competition between the bands with the opposite helical states, since the second band is not considered. 
In Fig. \ref{eta_dsdf} we plot the diode quality factor as a function of the S and F parts widths, which are $d_s$ and $d_f$ respectively.  This plot has important features that characterize the considered system.  From the figure we can recognize the sharp transition from $\eta =0$ to a nonzero value. This rapid change corresponds to the transition between normal and superconducting state and can be characterized by the critical width of the S part $d_s^{crit}$ .  It can be seen from the figure that there is an optimal value of $d_s$ and $d_f$ to reach the highest quality factor $\eta$ .  For larger $d_s$ and $d_f$ the quality factor will gradually decline since the characteristic lengths of the superconducting correlations in the S and F parts become smaller than the geometrical sizes of the hybrid structure.

\begin{figure}[t]
	\includegraphics[width=\columnwidth,keepaspectratio]{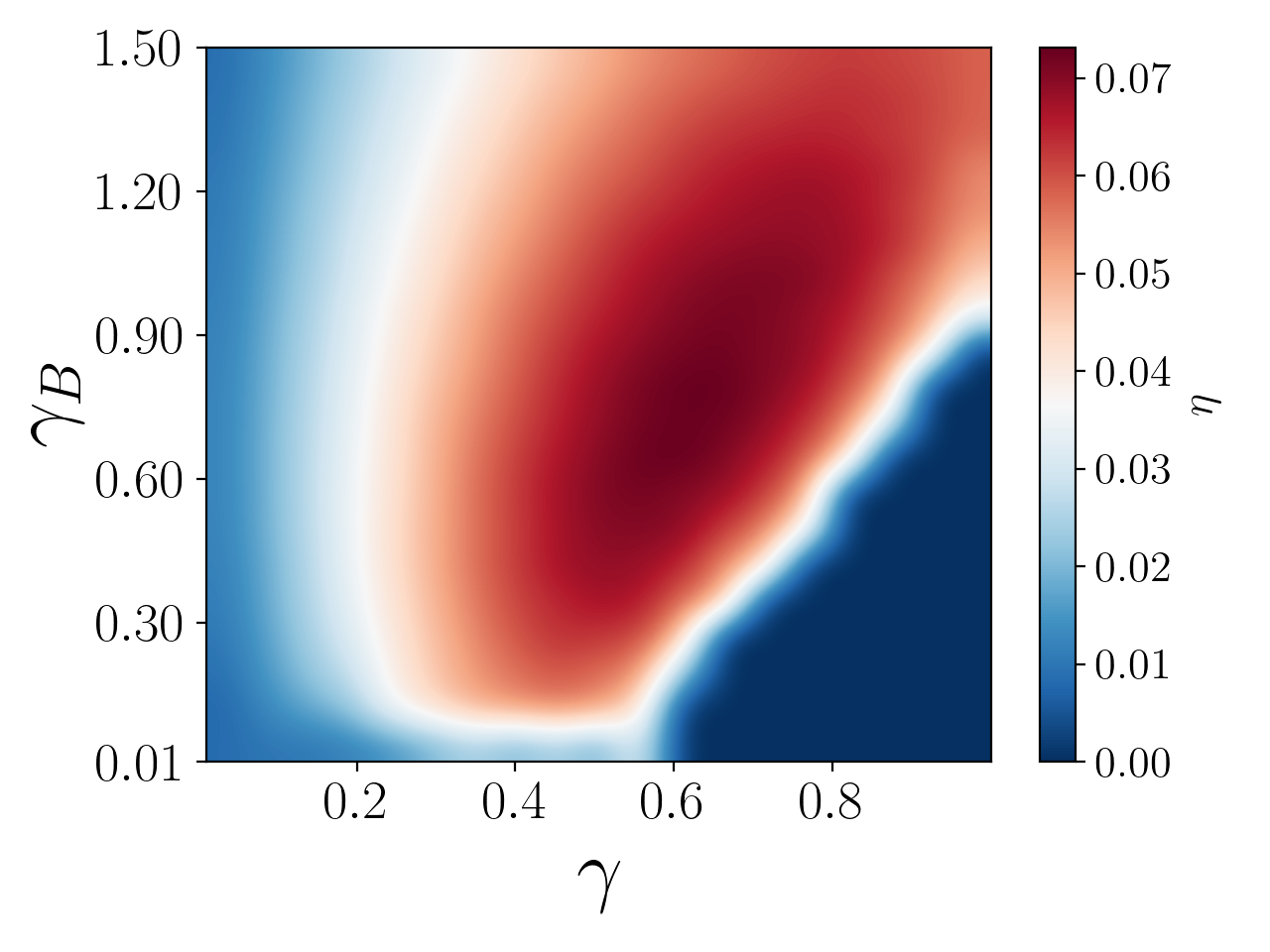}
	\caption{ Phase diagram of the diode efficiency $\eta$  as a function of the interface parameters $\gamma -\gamma_B$. The parameters of the calculation: $T= 0.1T_{cs}, d_s=1.2 \xi, d_f = \xi, H = 0.3$
	}
	\label{eta_gamma}
\end{figure}

Another important aspect of this study is to investigate the parameters that control the proximity effect. In Fig.~\ref{eta_gamma}  $\gamma - \gamma_B$ diagram for $\eta$ is illustrated. Interestingly, one can notice that there is an optimal nonzero value of the interface transparency parameter $\gamma_B$, i. e. having a perfect transparency does not result in a better diode efficiency. This can be explained by the following argument. When the transparency of the interface is low ($\gamma_B>1$) the proximity effect is strongly suppressed, the mutual coupling between S and F parts of the structure becomes weaker. Thus, the SDE reduces as $\gamma_B$ keeps increasing. On the other hand, when $\gamma_B<1$ there can be one of the two cases depending on the value of $\gamma$. In the case when $\gamma>0.5$ the inverse proximity effect becomes too strong and the superconducting state in the S part is suppressed. Hence, this results in the vanishing of the SDE. When $\gamma<0.5$ the superconductivity does not vanish, but the diode effect is small at $\gamma_B \approx 0$ and reaches its maximum value at finite $\gamma_B$ . From this result we conclude that for the highest $\eta$ it is optimal to have a finite transparency parameter across the S/F interface. 

\section{Discussion and Conclusion}\label{Fin}
In this work we have investigated the details of the superconducting diode effect in the S/F/TI hybrid structure. Employing the quasiclassical formalism of the Usadel equations we have introduced a simple model of the S/F/TI structure to study the diode quality factor $\eta$ as a function of various parameters of the structure. For this purpose we have examined the phase diagrams of $\eta$ revealing the most favourable conditions for the SDE in the system. From $H-T$ diagram we have found that the highest diode quality factor is achieved at lower temperatures and at a specific $H$. From the same diagram we have recognized the critical field of the superconducting system. We have found the critical width $d_s^{crit}$  on  $d_s -d_f$ diagram that corresponds to the minimal $d_s$ for the system to be in a superconducting state. Finally we have analyzed the diode quality factor as a function of the interface parameters. It has been shown that to reach the highest $\eta$ it is optimal to have finite transparency of the interface. These findings may help designing and developing the SDE devices based on the proximity effect.

\section{Acknowledgements}
The formulation of the model and the calculation of Figure \ref{eta_th} were supported by Russian Science Foundation Project No. 23-72-30004. The calculation of Figure \ref{eta_dsdf}  was supported by the Foundation for the Advancement of Theoretical Physics and Mathematics “BASIS” grant number 22-1-5-105-1. The calculation of Figure \ref{eta_gamma} was supported by theMirror Laboratories Project and the Basic Research Program of the HSE University

\section*{References}
\bibliographystyle{iopart-num-nourl}
\bibliography{diode}
\end{document}